\documentclass[prb,twocolumn,superscriptaddress,amsmath,amssymb,aps,nofootinbib]{revtex4-2}
\usepackage{mathbbol,amsmath,amsfonts,amssymb,bm,dsfont,mathtools}
\usepackage{graphicx,psfrag,array,mathtools,lipsum,mathrsfs}
\usepackage{adjustbox}
\usepackage{siunitx}
\usepackage{braket}
\usepackage{comment}
\usepackage{physics}
\usepackage[dvipsnames]{xcolor}

\def\bk{{\mathbf{k}}}
\def\br{{\mathbf{r}}}
\def\bxi{{\boldsymbol{\xi}}}
\def\bU{{\mathbf{U}}}

\providecommand{\ignore}[1]{}
\providecommand{\aucmnt}[1]{#1}
\renewcommand{\aucmnt}[1]{}

\newcommand{\Comment}[1]{}

\begin{document}

\title{Entangled-Beam Reflectometry and Goos-H\"anchen Shift}

\author{Q. Le Thien}
\affiliation{Department of Physics, Indiana University, Bloomington,
IN 47405, USA}
\affiliation{Institute for Quantum Computing, University of Waterloo, Waterloo, N2L 3G1, ON, Canada}

\author{R. Pynn}
\affiliation{Department of Physics, Indiana University, Bloomington,
IN 47405, USA}

\author{G. Ortiz}
\email{ortizg@iu.edu}
\affiliation{Department of Physics, Indiana University, Bloomington,
IN 47405, USA}
\affiliation{Institute for Quantum Computing, University of Waterloo, Waterloo, N2L 3G1, ON, Canada}
\affiliation{Institute for Advanced Study, Princeton, NJ 08540, USA}

\begin{abstract}
We introduce the technique of {\it Entangled-Beam Reflectometry} for extracting spatially correlated (magnetic or non-magnetic) information from material surfaces or thin films. Our amplitude- and phase-sensitive technique exploits the coherent nature of an incoming entangled probe beam, of matter or light waves, undergoing reflection from the surface. Such reflection encodes the surface spatial structure into the probe's geometric and phase-derived Goos-H\"anchen shifts, which can then be measured to unveil the structure. We investigate the way these shifts depend on the wave packet widths, and illustrate our technique in the case of in-plane periodic (non-)magnetic structures by utilizing spin-path mode-entangled neutron beams. 
\end{abstract}

\maketitle













{\it Introduction.---}
Advances in quantum materials with complex inhomogeneous 
structures are often thwarted by the need for tools to unravel them. Reflectometry is a non-destructive general technique to identify these structured materials that utilizes some form of light \cite{xray} or matter waves, e.g., neutrons \cite{Majkrzak2003,Fitzsimmons,Temst}, as test probes. With a broad range of modalities, such as non-polarized, polarized, specular and off-specular, methods like neutron vector magnetometry and soft resonant x-ray reflection can reveal structural features of layered magnetic materials. They are, however, limited to uncovering either statistically-averaged electromagnetic quantities or variations along particular spatial directions. Then, a natural line of inquiry is whether exploiting entanglement of the probe overcomes those limitations and, ultimately, provides additional structural information. Here, we propose to exploit the {\it entangled Goos-H\"anchen} (GH)  effect \cite{newton,gh1,gh2} as a novel quantum sensing strategy. 

\begin{figure}[htb]
\centering
\includegraphics[width=0.45\textwidth]{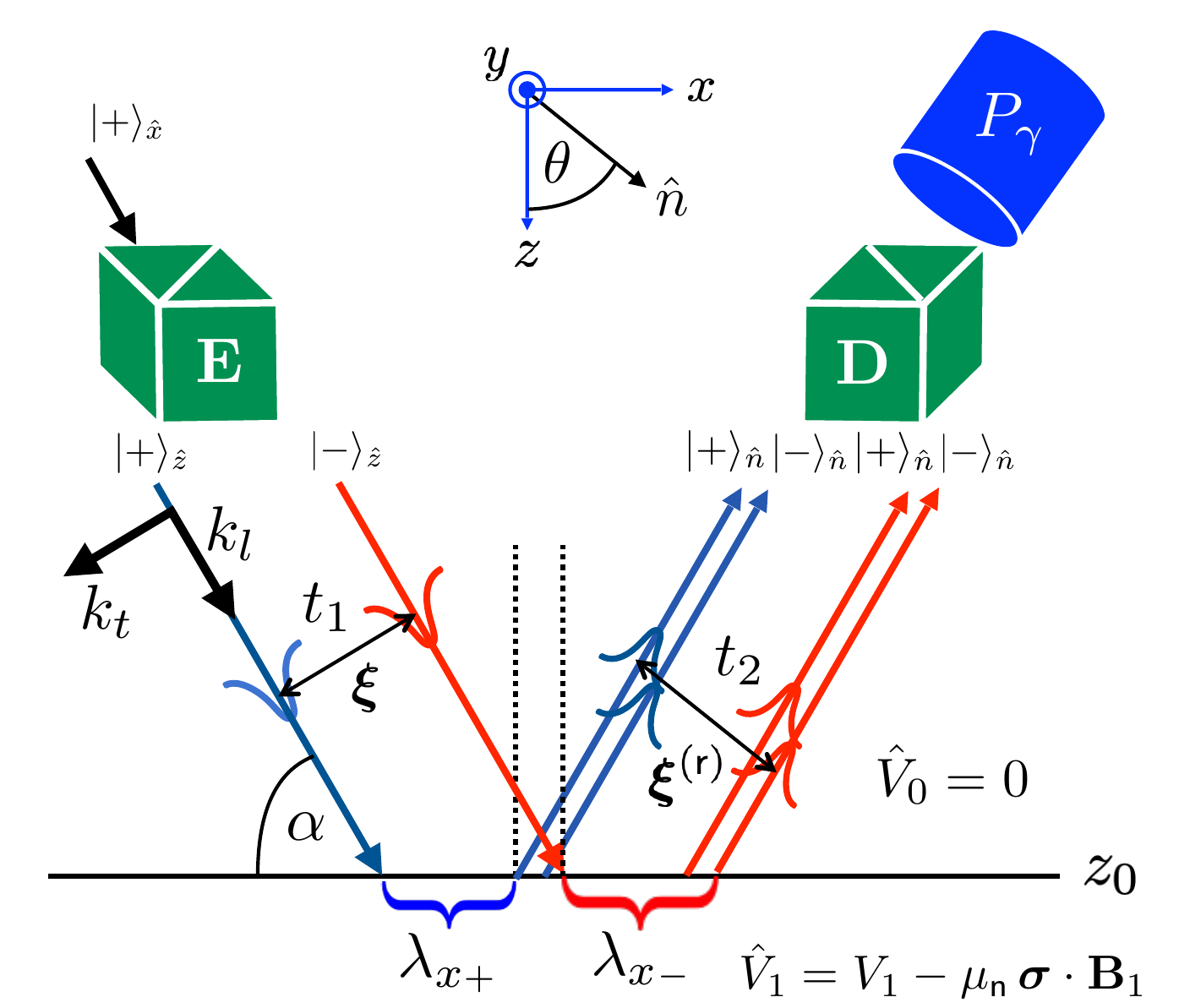}
    \caption{Setup for measuring the entangled GH-shift. The incoming beam traverses an {\it entangler} ({\bf E}) before impinging on the surface. The reflected entangled state may be subject to a {\it disentagler} ({\bf D}) before one extracts the shift from the measured polarization $P_\gamma$. The entanglement vector $\bxi$ is arbitrary and may even be purely longitudinal. The sample quantization axis is $\hat n$ (${\bf B}_1=B_1 \hat n$) with eigenspinors $\ket{\pm}_{\hat n}$. }
    \label{setup-entangled}
\end{figure}

The GH-shift refers to the in-plane lateral shift, $\lambda_x$, of the center of a beam of waves undergoing total reflection (TR) from a surface (along the $x$-axis in Fig.~\ref{setup-entangled}). This phenomenon (disputed to have been) suggested in light by Sir Isaac Newton \cite{newton}, was measured by Goos and H\"anchen in 1947 \cite{gh1,gh2}. The first theoretical description of the phenomenon focused on the phase shift $\Phi(k_{0z})$ of an incoming plane wave of momentum $\bk_0=(k_{0x},k_{0y},k_{0z})$, 
%
$\lambda_x^{\mathrm{ACH}} \equiv \frac{k_{0x}}{k_{0z}} \frac{d\Phi(k_{0z})}{dk_{0z}}$,
%
an expression known as the Artmann-Carter-Hora (ACH) formula \cite{artmann,carterhora}.  The connection between this and the geometric spatial shift $\lambda_x$ \cite{Frank14} was made by arguing that $\lambda_x^{\mathrm{ACH}}$ could be expressed as the effective displacement resulting from the group delay time $\tau$ the wave experiences between arrival and departure from the surface, first noted by Agud\'in \cite{Agudin68},
$    \lambda_x^{\rm ACH} = \frac{k_{0x}}{m} \frac{d\Phi(k_{0z})}{dE_z} = v_x \tau \equiv \lambda_x$, 
where $2m E_z = k_{0z}^2$ and $m v_x=k_{0x}$, assuming non-relativistic particles of mass $m$ (and reduced Planck constant $\hbar=1$).  Since these original contributions various attempts at deriving such a geometric shift using wave packets have been advanced in the literature \cite{renard64,lotsch68,li07,Ignatovich04}. Recently, de Haan {\it et al.} \cite{neutronGH} claimed observation of the effect in neutron optics, but consensus remains elusive \cite{dehaanResponse}. 

In the present work we introduce the general technique of {\it Entangled-beam (matter or light waves) Reflectometry} which leverages the GH-shift that results after the incoming beam,  entangled by design in various ways, is reflected by a magnetic (or birefringent) surface slab. In conjunction with Wollaston prisms or radio-frequency flippers, the GH effect can be exploited to generate arbitrary mode-entangled neutron beams \cite{contextuality}. We show how entanglement of the incoming neutron beam can be used to extract information about inhomogeneous (magnetic and/or non-magnetic) structures of the probed surface. The entangled beam is a very sensitive probe of magnetism along the direction perpendicular to the surface; in particular, it is very sensitive to thin magnetic layers.  
 
{\it Wave Packet Width and the Goos-H\"anchen Shift.---}
Why do the GH-shift expressions seem  independent of the incoming wave packet width? After all, a geometric GH-shift has a physical meaning only for finite width wave packets, making the identification $\lambda_x=\lambda_x^{\rm ACH}$ questionable. We next illustrate by means of an elliptical wave packet analysis that this is indeed the case. Our motivation is grounded on available experimental evidence where longitudinal, $\Delta_l$, and transverse, $\Delta_{t1}$ and $\Delta_{t2}$, coherence lengths might differ \cite{Kaiser,TransverseCoherence,PushinCoherenceLength}. 

Consider an incoming wave packet in momentum space centered at $\br_c=(x_c,y_c,z_c)$, (${\cal N}$ is a normalization factor)
\begin{eqnarray}\hspace*{-0.5cm}
\Psi^{({\sf i})}(\bk)={\cal N} \, e^{-\frac{\Delta_x^2}{2}(k_x-{k}_{0x})^2 -\frac{\Delta_y^2}{2}k_y^2
-\frac{\Delta_z^2}{2}(k_z-\bar{k}_{0z})^2-i \bk\cdot \br_c } ,
\label{momentum wf exact}
\end{eqnarray}
where we omit spinor notation for now, and $k_{0y}=0$. Here, 
$\Delta_x \Delta_z = \Delta_l \Delta_{t1}$, $\Delta_z = \sqrt{\Delta_l^2 \sin ^2\alpha +\Delta_{t1}^2 \cos ^2\alpha}$,  and $\Delta_y=\Delta_{t2}$ since, without loss of generality, we have chosen the plane of incidence to be the $xz$-plane (Fig. \ref{setup-entangled}). The
Gaussian peaks at
$    \bar{k}_{0z} = \frac{\sin \alpha 
   \left( k_{0l}\, \Delta_l^2-k_{x} \cos \alpha \,
   \left(\Delta_l^2-\Delta_{t1}^2\right)\right)}{\Delta_z^2}$ along $k_z$ (Momenta along different frame axes are related by $k_x=k_l \cos \alpha$, $k_z=k_l \sin \alpha$). 
This wave packet gets reflected by a surface whose reflection coefficient $R(k_z)$ depends only on the momentum component normal to the surface because of translation symmetry along the $x$ and $y$ directions. For TR, the resulting reflected wave packet (we assume no absorption)
can be expressed in terms of a momentum-dependent phase shift, ${\Phi}(k_z) \in \mathbb{R}$, 
$\Psi^{({\sf r})}(\bk)= R(k_z)\Psi^{({\sf i})}(\bk) =e^{i {\Phi(k_z)}} \, \Psi^{({\sf i})}(\bk)$. 
In coordinate representation, the time($t$)-dependent reflected wave function, 
\begin{eqnarray}
\label{elliptical spatial outgoing integral}
\Psi^{({\sf r})}(\br,t)= \frac{1}{(2\pi)^{\frac{3}{2}}}
\int d \bk \ \Psi^{({\sf r})}(\bk) \ e^{i (\bk^{(\sf r)} \cdot \br - \frac{k^{(\sf r)2}}{2 m} t )} ,
\end{eqnarray}
carries information about the surface, where $\alpha<\alpha_c$  (critical angle), and we define $\bk^{(\sf r)} = (k_x,k_y,-k_z)$. 


To extract the GH-shift one needs to examine the spatial wave function. Unfortunately, the integral in Eq.~\eqref{elliptical spatial outgoing integral} does not admit a closed analytic form. To proceed, 
we assume $\Delta_z$ to be larger than the semi-classical penetration depth 
$\delta_{\sf p}(k_{z})= -\frac{i}{2}  \frac{d \ln R(k_z)}{d k_z}$ 
\cite{Ignatovich04},
and expand ${\Phi}$ (up to second order in $k_z$) about $\bar k_{0z}$
\begin{eqnarray}
\label{ln r expansion}
\hspace*{-0.5cm}
\Phi(k_z)= -i \ln \bar R +
2\bar \delta_{\sf p} \ (k_z-\bar{k}_{0z})+\overline{W}^2 \ (k_z-\bar{k}_{0z})^2 ,
\end{eqnarray}
where 
$W^2(k_{z}) = -\frac{i}{2} \frac{d^2 \ln R (k_z)}{d k_z^2}$, 
and $\bar R$, $\bar \delta_{\sf p}$, $\overline{W}^2$ are evaluated at $k_{z}=\bar k_{0z}$. While this expansion is sufficient for the spherical wave packet case ($\Delta_l=\Delta_{t1}=\Delta_{t2}$) \cite{Ignatovich04}, it is not in the elliptical case
because $\Psi^{({\sf r})}(\br,t)$ intertwines the $x$ and $z$ coordinates due to the $k_{x}$-dependence in $\bar k_{0z}$.  

Consider next Eq. \eqref{momentum wf exact} and assume that $k_x \approx k_{0x}$. Then, the 
quasi-spherical approximation amounts to
\begin{align}
    \left | \frac{\Delta_l^2-\Delta_{t1}^2}{\Delta_l^2} \right | \ll  \frac{k_{0l}}{k_{0x} \cos \alpha} = \frac{1}{\cos^2 \alpha} ,
\end{align}
and $\bar k_{0z}$ becomes 
\begin{align}
    \label{elliptical kz 0th}
    \bar k_{0z} \approx \tilde k_{0z} = \left(\frac{\Delta_l}{\Delta_z} \right)^2  k_{0l} \sin \alpha    = \left(\frac{\Delta_l}{\Delta_z} \right)^2 k_{0z},
\end{align}
where, to lowest-order, the ellipticity in the incoming wave packet simply renormalizes the peak value $\bar k_{0z}$ to $\tilde k_{0z}$. This replacement, $\bar k_{0z} \rightarrow\tilde k_{0z}$, extends to all other quantities which are dependent on $\bar k_{0z}$, such as $\widetilde \Phi$, $\tilde \delta_{\sf p}$ and $\widetilde W^2$. Using this quasi-spherical approximation, the integral in Eq.~(\ref{elliptical spatial outgoing integral}) now becomes separable:
%
$    \Psi^{({\sf r})}(\br,t) \approx \Psi^{({\sf i})}_x(x,t) \Psi^{({\sf i})}_y(y,t) \Psi^{({\sf r})}_z(z,t)$,
%
where $\Psi^{({\sf i})}_{x,y}$ represent the $xy$-components of the time-evolved incident wave packet of width $\Delta_{x,y}$ and mean momentum $(k_{0x},0)$, and $\Psi^{({\sf r})}_z(z,t)$  a modified Gaussian wave packet with  dispersion 
$\Delta_z(t) = \Delta_z \sqrt{1+\left(\frac{t-2 m \widetilde W^2 }{m\Delta_z^2} \right )^2}$  \cite{SM},
showing that $\widetilde W^2$ reduces the effective dispersion time of the wave packet. 
The reflected wave packet centers about $\br^{(\sf r)}(t)=(r^{({\sf r})}_x(t),r^{({\sf r})}_y(t),r^{({\sf r})}_z(t))=(0,0,2\tilde\delta_{\sf p})+ \br^{(\sf r)}_c +  \frac{\tilde \bk_{0}^{(\sf r)}}{m} t$, where $\br^{(\sf r)}_c = (x_c,y_c, -z_c)$ and $\tilde\bk^{(\sf r)}_{0} = (k_{0x}, 0, -\tilde k_{0z}) $.

%
%
%
%

We are now in a position to address our original question. From the time-evolved incoming state  one can determine the moment at which the center of the wave packet reaches the surface. The time interval between the arrival of $\Psi^{(\sf i)}$'s center and the departure of $\Psi^{(\sf r)}$'s center from the surface, $\tau_{\sf WP}$ (the wave packet analog of the group delay time $\tau$), can be determined from $\Psi^{(\sf r)}_z(z,t)$. The geometric GH-shift $\lambda_x$ can then be defined as the displacement of $\Psi^{(\sf i)}_x(x,t)$'s center during the interval $\tau_{\sf WP}$  
\begin{eqnarray}
    \label{gh shift elliptical}
    \lambda_x = v_x \tau_{\sf WP} = 2 \tilde \delta_{\sf p}  \cot \alpha \left(\frac{\Delta_z}{ \Delta_l}\right)^2,
\end{eqnarray}
which shows a wave packet width dependence. Note that the second-order term,  $\widetilde{W}^2$, does not affect the GH-shift. 

While Eq.~\eqref{gh shift elliptical} addresses our initial puzzle, a natural question arises whether the ACH formula still gives the same result as  Eq.~\eqref{gh shift elliptical}. The asymptotic reflection phase-shift, in the far-field limit ($t/m \gg \Delta_z^2, \widetilde W^2$), is $\widetilde{\Phi}(\tilde k_{0z})$ \cite{SM}.
%
%
Note that this expression is different from the reflection phase shift $\Phi(k_{0z})$ of an incoming plane wave. 
By analogy to the plane wave case 
one can define the wave packet version of the ACH formula
%
\begin{eqnarray} \hspace*{-0.6cm}
    \label{ACH formula correction}
    \lambda_x^{\mathrm{ACH}} &\equiv& \frac{k_{0x}}{k_{0z}} \frac{d {\widetilde\Phi(\tilde k_{0z})}}{d k_{0z}} = 2 \tilde\delta_{\sf p} \cot \alpha \left( \frac{\Delta_l}{\Delta_z} \right)^2,
\end{eqnarray}
%
which, in general, is different from the geometric shift in  Eq.~\eqref{gh shift elliptical}. For a spherical wave packet the width dependence disappears and thus Eq.~\eqref{gh shift elliptical} agrees with Eq.~\eqref{ACH formula correction}. 


{\it The Entangled Goos-H\"anchen Effect.---} We now turn to examine the GH-shift for an incoming mode-entangled, in spin (or polarization) and path, state \cite{Ashik21,contextuality}
\begin{eqnarray}
\label{momentum wf entangled}
\Psi^{({\sf i})}_\bxi (\bk)= \Psi^{({\sf i})}(\bk) \otimes \boldsymbol{\chi}^{({\sf i})}_{\bk \cdot \bxi} \ ,
\end{eqnarray}
where the two-component spinor $\boldsymbol{\chi}^{({\sf i})}_{\bk \cdot \bxi}=\frac{1}{\sqrt{2}}\binom{e^{ - i \bk \cdot \bxi / 2}}{e^{ i \bk \cdot \bxi / 2}}$, written in the $\sigma^z$-basis ($\ket{+}_{\hat z}$ and $\ket{-}_{\hat z}$), carries information about the displacement vector $\bxi$ separating the centers of the two incoming coherent wave packets (Fig. \ref{setup-entangled}). Upon TR from a magnetic (or birefringent) sample, the reflection operator $\hat R(k_z) = \mathrm{diag} (R_+,R_-)$, in its orthonormal eigenbasis $\ket{\pm}_{\hat n}$, defines an effective quantization axis $\hat{n} = (\sin\theta \cos\varphi, \sin\theta \sin\varphi, \cos \theta)$, where $\theta$ and $\varphi$ are the usual Bloch sphere angles \cite{SM}. The two bases are related by a unitary unimodular rotation $\bU(\theta, \varphi)$: $(\ket{+}_{\hat n} \ \ket{-}_{\hat n})^{\rm T}= \bU^{\rm T}(\theta,\varphi) (\ket{+}_z \ \ket{-}_z)^{\rm T} $, where  $\bU=\begin{pmatrix} \cos \frac{\theta}{2} & -e^{- i \varphi} \sin \frac{\theta}{2}\\ e^{i \varphi} \sin \frac{\theta}{2} & \cos \frac{\theta}{2}\end{pmatrix}$ and T denotes matrix transposition.
%
%

 Hence, the reflected wave function is given by
\begin{eqnarray}\hspace*{-0.6cm}
    \label{outgoing wf entangled exact}
    \Psi^{({\sf r})}_\bxi (\br,t)\! = \! \frac{1}{(2\pi)^{\frac{3}{2}}} \! \int \!\! d\bk \, \hat{R}(k_z) \bU^\dagger \Psi^{({\sf i})}_\bxi (\bk)  \, e^{i (\bk^{(\sf r)} \cdot \br - \frac{k^{(\sf r)2}}{2 m} t )} 
\end{eqnarray}
in the surface $\ket{\pm}_{\hat n}$-basis, and in the quasi-spherical limit can be expressed as
\begin{eqnarray}\hspace*{-0.3cm}
    \label{elliptical spatial outgoing 0th entangled}
   \Psi^{(\sf r)}_\bxi(\br,t) 
 &=& \sum_{\mu,\nu=\pm} \bU^*_{\mu\nu} \Psi^{(\sf r)}_{\mu\nu} (\br,t) \otimes \ket{\nu}_{\hat n} ,
\end{eqnarray}
where $\Psi^{(\sf r)}_{\mu\nu}(\br,t)$ has its center at $\br_{\mu\nu}^{(\sf r)}(t)$ \cite{SM}.
Here, indices $\mu$ and $\nu$ carry physical meanings: $\mu$ is related to the incoming spin quantization axis $\hat{\sf e}$ ($=\hat z$ in Eq. \eqref{momentum wf entangled}), and $\nu$ is associated to the reflected one. Hence, upon TR, an incoming $\mu$ spin state is split into $\Psi^{(\sf r)}_{\mu+}$ and $\Psi^{(\sf r)}_{\mu-}$, with centers separated by $2(\tilde \delta_{{\sf p}+} - \tilde \delta_{{\sf p}-} )\hat{z}$ due to experiencing different penetraton depths $\tilde \delta_{{\sf p}\nu}$ and GH-shifts.  

Two physical effects emerge from Eq.~\eqref{elliptical spatial outgoing 0th entangled}. The first concerns the generated entanglement pattern. Generically, the reflected state splits coherently into four wave packets (Fig. \ref{setup-entangled}), yet it remains mode-entangled with respect to the distinguishable  path and spin subsystems decomposition. Secondly, if the incoming spin quantization axis $\hat{\sf e}$ coincides with that of the sample $\hat n$, there will only be two reflected wave packets $\Psi^{({\sf r})}_{++}$ and $\Psi^{({\sf r})}_{--}$, separated by a reflected entanglement vector $\bxi^{(\sf r)} = \br_{++}(t) - \br_{--}(t)$, with
%
$\xi_z^{({\sf r})} = -\xi_z+2(\tilde \delta_{{\sf p}+} - \tilde \delta_{{\sf p}-}) , \bxi_\|^{({\sf r})} =\bxi_\|=(\xi_x,\xi_y)$.
%
Hence, the distance between the two wave packet centers, i.e., the entanglement length \cite{Ashik21}, can be increased or decreased depending on the difference between penetration depths $\tilde \delta_{{\sf p}\pm}$. Furthermore, in this particular case one can identify a simple physical picture of the entangled GH-shift, i.e., each incoming spin state $\ket{\pm}_{\hat z}$ experiences the GH-shift $\lambda_{x\pm}$ shown in Fig.~\ref{setup-entangled}.

{\it Measuring the Entangled Goos-H\"anchen Shift.}
We next describe a way to extract the GH-shift from polarization measurements performed at the far-field as in Fig.~\ref{setup-entangled}. The (asymptotic) polarization components, $P_\gamma(\bxi)=\lim_{t\rightarrow\infty}P_\gamma(\bxi,t)$, $\gamma=x,y,z$, can be expressed as 
\begin{eqnarray}\hspace*{-0.5cm}
   P_\gamma(\bxi) &=&  \sum_{\mu,\mu',\nu,\nu'=\pm}  \!\! \bU^{\;}_{\mu\nu} \bU^*_{\mu'\nu'}  {\;}_{\hat n}\! \bra{\nu}\frac{\sigma^\gamma}{2} \ket{\nu'}_{\hat n} C^{\mu\mu'\nu\nu'} ,
    \label{polarization master}    
\end{eqnarray}
with $C^{\mu\mu'\nu\nu'}=\lim_{t\rightarrow\infty} \int d\br \ \Psi^{(\sf r)*}_{\mu\nu} (\br, t) \Psi^{(\sf r)}_{\mu'\nu'} (\br, t)=\Lambda_{\mu\mu'\nu\nu'} \, e^{i \eta_{\mu\mu'\nu\nu'}}$, 
%
%
where $\Lambda_{\mu\mu'\nu\nu'}\le 1$ encodes the overlap between reflected spin states at the detector \cite{SM}, and the asymptotic reflected phase difference, $\eta_{\mu\mu'\nu\nu'}\in \mathbb{R}$, is given by ($\Delta \widetilde \Phi_{\nu\nu'}=\widetilde \Phi_\nu - \widetilde \Phi_{\nu'}$)
\begin{eqnarray}
    \eta_{\mu\mu'\nu\nu'} &=& \Delta \widetilde \Phi_{\nu\nu'} + \frac{\mu-\mu'}{2} \widetilde\bk_0 \cdot \bxi \ . 
    \label{outgoing phase diff}
\end{eqnarray}
%
%
One can then use the generalized ACH formula,
\begin{eqnarray}
\lambda^{\rm ACH}_{x+} - \lambda^{\rm ACH}_{x-} &\equiv& \frac{k_{0x}}{k_{0z}} \frac{d {\Delta \widetilde \Phi_{\nu\nu'}(\tilde k_{0z})}}{d k_{0z}} ,
\label{eq:GACH}
\end{eqnarray}
and from the measured phase shift  $\Delta \widetilde \Phi_{\nu\nu'}(\tilde k_{0z})$ determine the relative GH-shift \cite{SM}. For perfectly collimated beams, had one chosen $\bxi$ along the $y$-axis one would have eliminated the last term in Eq. \eqref{outgoing phase diff}.

Using a disentangler before the detector (Fig. \ref{setup-entangled}) affects the $\bxi$-dependence of the measured polarization. The ultimate goal of this resulting {\it spin-echo entangled-beam} technique is to make sure that the observed phase shift is due only to the reflection process. For instance, if one chooses  $\hat{\sf e} = \hat n=\hat{\sf d}$, $\eta_{\mu\mu'\nu\nu'} \rightarrow  \Delta \widetilde \Phi_{\nu\nu'}$ \cite{SM}.


{\it Entangled-Beam Reflectometry.---} 
%
%
In entangled-beam reflectometry three length scales enter into the problem, the beam diameter, the wave packet width $\Delta_{x(y)}$, and the surface-projected entanglement length $\xi_\|$; we assume the beam diameter is the largest length scale. Consider first the case where the slab structural variations happen along the $z$-axis. There are practical advantages in using entangled beams in TR mode as opposed to traditional reflectometry techniques \cite{Fitzsimmons}. For instance, if the sample is non-magnetic or its effective magnetic quantization axis satisfies $\hat n = \hat{\sf e} = \hat{\sf d}$, the observed spin-echo polarization (Fig.~\ref{setup-entangled}) measures the spinor's reflection phase shift difference $\Delta\tilde\Phi_{+-}$ directly \cite{SM}. If that is not the case, there is a remnant $\bxi$-dependence in the measured polarization \cite{SM}. Nonetheless, under particular experimental conditions (e.g., entangled thermal neutron beams), because of an averaging over the incoming beam divergence, one may eliminate that $\bxi$-dependence (an example is presented in \cite{SM}). In this latter case, the main advantage of our technique over the standard polarized-beam reflectometry is in sensitivity,  because the spin-echo polarization shows strong signals at larger values of the reflectivity than conventional spin asymmetry measurements. This increased sensitivity is more dramatic in the case of thin magnetic multilayers.

We next consider the novel application of our technique to probe in-plane ($xy$-plane in Fig. \ref{setup-entangled}) inhomogeneities. Now, the characteristic period $p$ of the surface inhomogeneities competes with the beam length scales. 
If the distribution of nuclear density or electromagnetic fields on the surface is inhomogeneous,  i.e., $\hat{V}(\br)$, then, each spinor component experiences a different potential and, consequently, a different reflection operator $\hat{R}(\br_{\|},k_z)$ where $\br_\|=(x,y)$. (Note that the sample material is three-dimensional and structural variations along $z$ are encoded in the $k_z$-dependence of $\hat{R}(\br_\|,k_z)$.) Does our entangled probe provide additional information? We show next that this is indeed the case. 

%
%

If only the nuclear density varies (no magnetic fields),  $\hat{R}(\br_\|,k_z) = R(\br_\|, k_z) \, \mathbb{1}$, the observed spin-echo beam polarization $\overline{P}_x(\bxi_\|,t)$ ($\xi_z=0$),  averaged over incoming impact parameters (spanning the area $A$), is \cite{SM}
\begin{eqnarray}
    \overline{P}_x(\bxi_\|,t) =\frac{1}{A}  \mathfrak{R} \big[ & \int_A dx_c d y_c \int d\br_\| dk_z \  R^*(\br_+,k_z) R(\br_-,k_z)   \nonumber \\ 
    & \times |\Psi^{(\sf i)} (\br_\|,k_z,t)|^2 \big] ,
    \label{grating-reflection}
\end{eqnarray}  
where $\mathfrak{R}$ stands for the real part, $\br_\pm=\br_\| \pm \bxi_\|/2$,  and $\Psi^{(\sf i)}(\br_\|,k_z,t) = \int \frac{d\bk_\|}{2\pi} e^{i(\bk_\| \cdot \br_\|-\frac{k^{(\sf r)2}}{2m}t)} \Psi^{(\sf i)}(\bk)$. Eq. \eqref{grating-reflection} can be seen as the reflection analog of a purely transmission phase diffraction grating setup, where the role of the grating transmission function is played by 
the modulated surface reflection coefficient $R(\br_\|,k_z)$ \cite{quantumtale}. This also explains why the relationship between $\xi_\|$, $\Delta_{x(y)}$ and $p$ is not important for this non-magnetic case. Spin-echo resolved experiments from diffraction gratings have been done in the past \cite{Ashkar} but have been interpreted using a semi-classical theory \cite{quantumtale} rather than Eq. \eqref{grating-reflection}.  
\begin{figure}[htb]
\centering
\includegraphics[width=0.47\textwidth]{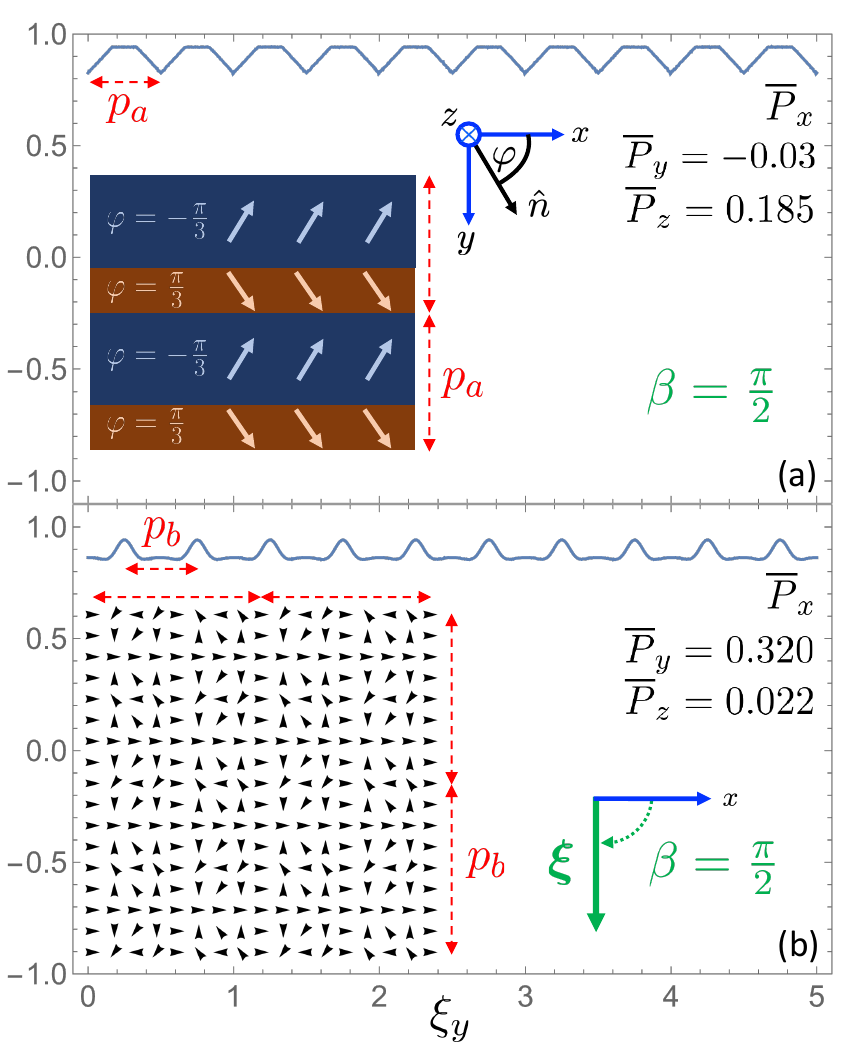}
    \caption{Average beam polarization $\overline P_x$ as a function of the entanglement length $\xi_y$ for two different magnetic structures with periods $p_a$ and $p_b$  ($\overline P_y$, $\overline P_z$ are $\xi_y$-independent constants). Insets show in-plane $\hat n(\br_\|)$ configurations: (a) 
    $(\frac{\pi}{2},\varphi)$, with 
    $\varphi=-\pi/3$ if $y \ {\rm mod} \ p_a  < \frac{2}{3} p_a$, and $\varphi=\pi/3$, otherwise. (b) $(\frac{\pi}{3},\pi \sin \frac{2\pi x}{p_b} \cos \frac{2\pi y}{p_b})$. One can vary the angle $\beta$, changing $\bxi$'s direction, to obtain structural correlation along different axes.}
    \label{setup-entangled-surface}
\end{figure}

The more challenging case presents when the in-plane inhomogeneities are magnetic. Our entangled-beam reflectometry technique is now applicable in the regime  $\Delta_{x(y)}< p \le \xi_\|$ because the incoming wave gets reflected in its locally-varied spinor basis $\hat n (\br_\|)$.
Here, the two spinor components effectively experience a different {\it local} reflection, $\hat{R}(r^{(\sf r)}_{\gamma+},k_z)$ and $\hat{R}(r_{\gamma- }^{(\sf r)} ,k_z)$, at their wave packet centers $r^{(\sf r)}_{ \gamma\pm}$ separated by $\bxi_\|$. The observed beam polarization $\overline{P}_\gamma(\bxi)$ can be determined by using Eq.~\eqref{polarization master} with these now $r^{(\sf r)}_{ \gamma\pm}$-dependent quantities, and then average over the impact parameters \cite{SM}. Furthermore, one can vary the direction of the vector $\bxi$ in $\Psi^{({\sf i})}_\bxi (\bk)$ to obtain spatial information of correlations along various directions on the surface. 
For perfectly collimated beams one may choose an entanglement vector $\bxi$  perpendicular to the incidence plane ($\bxi = \xi_y \hat{y}$ in  Fig.~\ref{setup-entangled-surface}) to cancel the $\bxi$-dependent phases in Eq.~\eqref{outgoing phase diff}, while we expect damping in the intensity in the far field limit. In many practical applications the beam divergence is non-negligible,  and one has to resort to {\it spin-echo measurements} to cancel the angular divergence of the beam as much as possible. However, in the magnetic case, spin-echo reflectometry \cite{SM} may not be straightforward since $\hat n (\br_\|)$ varies on the surface, and this is not known a priori.


As an illustration consider the magnetic structures, shown in Fig.~\ref{setup-entangled-surface}, which rest on semi-infinite slabs (Fig.~\ref{setup-entangled}). The optical potential $\hat{V}_1$ includes the nuclear potential and magnetic field ${\bf B}_1(\br_\|)=B_1 \, \hat{n}(\br_\|)$, where the magnitude $B_1$ is constant and the unit vector $\hat{n} (\br_\|)=(\theta, \varphi(\br_\|))$, with $\varphi(\br_\|)$ periodic. To increase the contrast in $\overline{P}_\gamma$, the grazing angle $\alpha=0.37^\circ$ is set close to the critical angle $\alpha_{c-}=0.43^\circ$ ($\alpha_{c+}=0.57^\circ$), corresponding to the $\ket{\nu}_{\hat n}=\ket{-}_{\hat n}$ state. Alternatively, one can use specifically-engineered magnetic layers with large in-plane difference in their reflection matrices $\hat R(\br_\|,k_z)$.  The entangled beam has $\bxi=\xi_y \hat{y}$. The structure in Fig.~\ref{setup-entangled-surface}(a) is periodic with  $\theta=\pi/2$ and $\varphi(\br_\|)=\begin{cases}
    -\pi/3 & \textrm{if}\ y \ {\rm mod} \ p_a < \frac{2}{3} p_a ,\\
    \pi/3 & \textrm{otherwise} .
\end{cases}$.
Interestingly, one can extract the period $p_a$ from 
$\overline{P}_x$. The asymmetry in the spatial extent of each magnetic-field region, within a period, is encoded in a non-vanishing value of $\overline{P}_y$. 
Inset (b) shows a two-dimensional periodic structure with $\theta=\frac{\pi}{3}$ and $\varphi(\br_\|)=\pi \sin \frac{2\pi x}{p_b} \cos \frac{2\pi y}{p_b} $. The period $p_b$ also shows up in $\overline{P}_x$, while a non-vanishing $\overline{P}_z$ signifies that ${\bf B}_1$ has an out-of-plane component. 


{\it Conclusions.---} 
The revealed {\it entangled Goos-H\"anchen effect} is a phenomenon that evinces the wave nature of light and matter, and the unique trait of quantum mechanics which Einstein epitomized as spooky. The intrinsic value of this  phenomenon rests on the possibility to exploit it by designing entangled probes that, after interaction with a target surface-distribution of electromagnetic fields, unveil by interference that surface information. This amplitude- and phase-sensitive technique the present work introduced is called  {\it Entangled-Beam Reflectometry and is not unique to a particular probe}. The information obtained depends on the nature of the probe and the {\it entanglement design}. For instance, neutron beams mode-entangled in spin and path provide information about inhomogeneous non-magnetic or magnetic-field surface distributions, by direct encoding in spinor phase differences. While off-specular diffuse scattering has been used to study non-magnetic structures \cite{sergis}, due to the low intensity of the partially-reflected beam its applicability is limited, a problem  we do not face in TR. By tuning the length and direction of the probe's entanglement vector, and in conjunction with spin-echo measurements, one can in principle map out the magnetism of the material's surface. Within an elliptical wave packet formulation, we proved the width dependence of the Goos-H\"anchen shift, and also clarified the distinction between geometric and ACH phase shifts \cite{artmann} calculations of the Goos-H\"anchen shift.

The method can also be applied when the reflectivity is less than unity and, since it directly yields the relative phase between neutron spin states, it potentially allows reflectivity data to be inverted to yield the scattering-length depth profiles \cite{Majkrzak2003}. Furthermore, the relative phase permits identification of thin magnetic layers that is more sensitive than traditional neutron spin asymmetry reflectometry because the measurements can be made at larger values of the reflectivity. It is our hope that this Entangled-Beam Reflectometry technique will be applied to unravel exotic electromagnetic structures, including characterization of topological materials.

{\bf Acknowledgments.}
We are indebted to conversations with David Baxter, Emilio Cobanera, Rebecca Dally, Jonathan Gaudet, Sam McKay, Charles Majkrzak, and Dmitry Pushin. The IU Quantum Science and Engineering Center is supported by the Office of the IU Bloomington Vice Provost for Research through its Emerging Areas of Research program. This research was undertaken thanks in part to funding from the Canada First Research Excellence Fund. GO gratefully acknowledges support from the Institute for Advanced Study.

\newpage

\onecolumngrid
\newpage 

\section*{Supplemental Material:\\
Entangled-Beam reflectometry and Goos-H\"anchen Shift}
\hspace*{5.5cm} Q. Le Thien, R. Pynn and G. Ortiz$^*$ 
\vspace*{0.5cm}

\twocolumngrid
In what follows, we expand on various aspects.

\section{Neutron optics with spin}
\label{appedix reflection spin}

\begin{figure}[htb]
\centering
\includegraphics[width=0.45\textwidth]{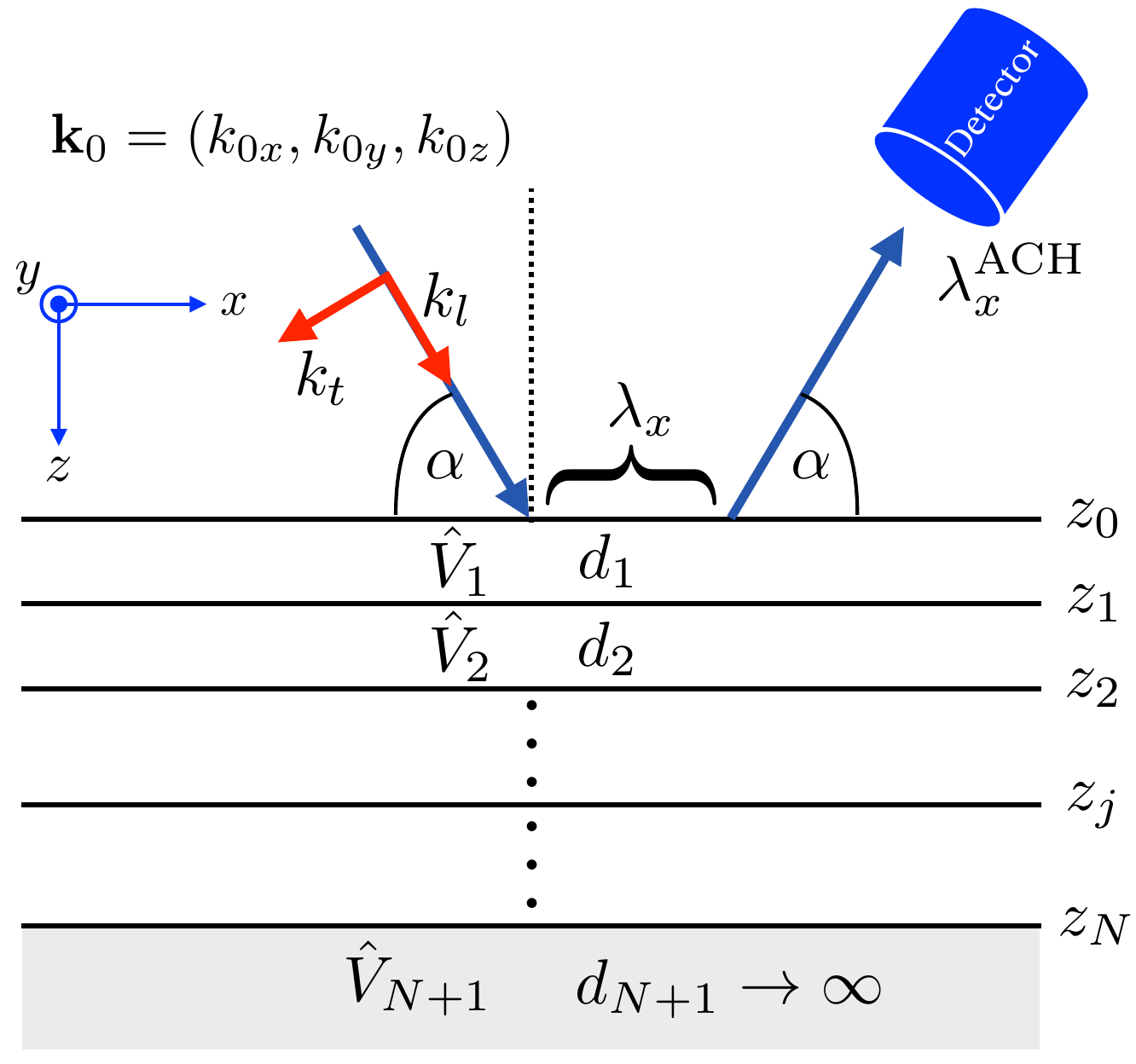}
\caption{A ray representation of the standard GH effect, assuming $\alpha\le\alpha_c$, the critical angle, above which the reflection is no longer total. An incoming wave of mean momentum ${\bf k}_0$ impinges on a surface at $z_0$ and gets reflected.
Geometric, $\lambda_x$, and phase-derived, $\lambda_x^{\rm ACH}$, GH shifts are indicated. Allowed magnetic structures along the $z$-axis are simply characterized by a stratified medium with homogeneous layers (of width $d_j$ and optical potential $\hat V_j$) labeled by the index $j=1,2,\cdots,N+1$. 
Momenta along different frame axes are related by $k_x=k_l \cos \alpha$, $k_z=k_l \sin \alpha$. }
    \label{setup}
\end{figure}

In the following we are assuming the optical potential approximation, where the medium represents a constant (spin or polarization dependent) refractive index \cite{Blundell92,Pleshanov96}. The magnetic structures along the $z$-axis are simply characterized by a stratified medium with layers, labeled by the index $j=0,1,\cdots,N+1$, and potential energy given by
\begin{eqnarray}
    \hat V_j=V_j - \mu_{\sf n} \,\boldsymbol{\sigma}\cdot {\bf B}_j .
\end{eqnarray}
Here, $V_j$ is a scalar potential, $\mu_{\sf n}<0$ for neutrons,  $\boldsymbol{\sigma}=(\sigma_x,
\sigma_y,\sigma_z)$ are Pauli matrices and $\hat{\bf b}_j={\bf B}_j/|{\bf B}_j|$ is the magnetic field unit vector  (see Fig. \ref{setup}). 

The spinor state and its first derivative in layer $j$ are given by
\begin{eqnarray}
 \Psi_j(z)&=& e^{i {\sf k}_j z} \chi^+_j + e^{-i {\sf k}_j z} \chi^-_j \nonumber \\
 \Psi'_j(z)&=& i {\sf k}_j (e^{i {\sf k}_j z} \chi^+_j - e^{-i {\sf k}_j z} \chi^-_j), 
\end{eqnarray}
which can be compactly written as
\begin{eqnarray} \hspace*{-1cm}
 \begin{pmatrix} 
 \Psi_j(z) \\
 \Psi'_j(z)
 \end{pmatrix}
 \!=\! {\sf B}_{j}(z) \, 
\boldsymbol{\chi}_j =
 \begin{pmatrix} 
 e^{i {\sf k}_j z} & e^{-i {\sf k}_j z} \\
 i {\sf k}_j e^{i {\sf k}_j z} & -i {\sf k}_j e^{-i {\sf k}_j z} 
 \end{pmatrix} \boldsymbol{\chi}_j \ ,
\end{eqnarray}
where the momentum operator,
\begin{eqnarray}
{\sf k}_j= \frac{k^+_j+k^-_j}{2}+\frac{k^+_j-k^-_j}{2}\, \boldsymbol{\sigma} \cdot \hat{\bf b}_j,
\end{eqnarray}
has eigenvalues $k^\pm_j=\sqrt{2m (E_z-(V_j\mp \mu_{\sf n} |{\bf B}_j|))}$ ($\hbar=1$), and the (two-component) spinor 
\begin{eqnarray}
\boldsymbol{\chi}_j=  
\begin{pmatrix} 
 \chi^+_j \\
 \chi^-_j
 \end{pmatrix}    
\end{eqnarray}
is expressed in terms of unnormalized spin-1/2 states $\chi^\pm_j$. 

Using the transfer matrix
\begin{eqnarray}
 \begin{pmatrix} 
 \Psi_j(z_{j-1}) \\
 \Psi'_j(z_{j-1})
 \end{pmatrix}
 = {\sf M}_{j}  
 \begin{pmatrix} 
 \Psi_j(z_{j}) \\
 \Psi'_j(z_{j})
 \end{pmatrix} ,
\end{eqnarray}
one can connect the spinor state at coordinates $z_j$ and $z_{j-1}$.
From boundary conditions at $z_{j-1}$ and $z_{j}$ in layer $j$ 
\begin{eqnarray} \hspace*{-0.5cm}
\begin{pmatrix} 
 \Psi_j(z_{j-1}) \\
 \Psi'_j(z_{j-1})
 \end{pmatrix} &=& {\sf B}_{j-1}(z_{j-1})  
\boldsymbol{\chi}_{j-1}
= {\sf B}_{j}(z_{j-1})  
\boldsymbol{\chi}_{j} \\
\begin{pmatrix} 
 \Psi_j(z_{j}) \\
 \Psi'_j(z_{j})
 \end{pmatrix}
 &=& {\sf B}_{j}(z_{j})  
\boldsymbol{\chi}_{j} = {\sf B}_{j+1}(z_{j})  
\boldsymbol{\chi}_{j+1} ,
\end{eqnarray}
it results 
\begin{eqnarray}
 \begin{pmatrix} 
 \Psi_j(z_{j-1}) \\
 \Psi'_j(z_{j-1})
 \end{pmatrix}
 &=& {\sf B}_{j}(z_{j-1}) {\sf B}^{-1}_{j}(z_{j}) 
 \begin{pmatrix} 
 \Psi_j(z_{j}) \\
 \Psi'_j(z_{j})
 \end{pmatrix} \nonumber \\
 &=& {\sf M}_{j}  
 \begin{pmatrix} 
 \Psi_j(z_{j}) \\
 \Psi'_j(z_{j})
 \end{pmatrix} ,
\end{eqnarray}
with ($d_j=z_j-z_{j-1}$)
\begin{eqnarray} \hspace*{-0.5cm}
 {\sf B}^{-1}_{j}(z)  =\frac{1}{2}
 \begin{pmatrix} 
 e^{-i {\sf k}_j z} &  -i e^{-i {\sf k}_j z}{\sf k}^{-1}_j \\
 e^{i {\sf k}_j z} & i  e^{i {\sf k}_j z} {\sf k}^{-1}_j 
 \end{pmatrix},
 \end{eqnarray}
 and
\begin{eqnarray}
{\sf M}_{j} = 
 \begin{pmatrix} 
 \cos({\sf k}_j d_j) &  -\sin({\sf k}_j d_j){\sf k}^{-1}_j\\
 {\sf k}_j \sin({\sf k}_j d_j) &  \cos({\sf k}_j d_j)
 \end{pmatrix} ,
\end{eqnarray}
where the inverse of the momentum operator is
\begin{eqnarray}\hspace*{-0.5cm}
{\sf k}^{-1}_j= \frac{1}{k^+_j k^-_j}\left (\frac{k^+_j+k^-_j}{2}-\frac{k^+_j-k^-_j}{2}\, \boldsymbol{\sigma} \cdot \hat{\bf b}_j\right),
\end{eqnarray}

In the case where there are $N$ magnetic layers
\begin{eqnarray}
 \begin{pmatrix} 
 \Psi(z_{0}) \\
 \Psi'(z_{0})
 \end{pmatrix}
 = {\sf M}  
 \begin{pmatrix} 
 \Psi(z_{N}) \\
 \Psi'(z_{N})
 \end{pmatrix} ,
\end{eqnarray}
where ${\sf M}={\sf M}_1{\sf M}_2\cdots{\sf M}_N$ and, in general, $[{\sf M}_j,{\sf M}_{j'}]\neq 0$. They commute whenever ${\bf B}_j$ in different layers are collinear, i.e., $[\boldsymbol{\sigma} \cdot \hat{\bf b}_j ,\boldsymbol{\sigma} \cdot \hat{\bf b}_{j'}]=2i (\hat{\bf b}_{j}\wedge \hat{\bf b}_{j'})\cdot \boldsymbol{\sigma}=0$.

One can define reflection $\hat{R}$ and transmission $\hat{T}$ operators
\begin{eqnarray}
\Psi_R(z_0)&=& \hat{R} \, \Psi_0(z_0)= e^{-i {k}_{0z} z_0} \chi^-_0 ,\nonumber\\ 
\Psi_T(z_N)&=& \hat{T} \, \Psi_0(z_0)= e^{i {\sf k}_{N} z_N} \chi^+_N ,
\end{eqnarray}
by its action on $\Psi_0(z_0)=e^{i {k}_{0z} z_0}\chi^+_0$ with 
$\Psi(z_0)=\Psi_0(z_0)+\Psi_R(z_0)$ and $\Psi(z_N)=\Psi_T(z_N)$. These 
operators can be expressed in terms of the operator ${\sf M}$ as 
\begin{eqnarray}
\hat{T}&=& ({\sf M}_{11}+ i {\sf M}_{12} {\sf k}_N)^{-1} (\mathbb{1} +\hat{R}), \nonumber \\ 
\hat{R}&=&  ( k_{0z} + \boldsymbol{\kappa})^{-1}({k}_{0z} -  \boldsymbol{\kappa}),
\end{eqnarray}
where $\boldsymbol{\kappa}=-i({\sf M}_{21}+ i {\sf M}_{22} {\sf k}_{N+1})
({\sf M}_{11}+ i {\sf M}_{12} {\sf k}_{N+1})^{-1}$.

In the TR situation (without absorption) the operator $\hat R$ has to be unitary. That means that the basis that diagonalizes $\hat R$ must also diagonalize the operator $\boldsymbol{\kappa}$
\begin{eqnarray}
    \hat R = {\bf U}\, {\rm diag}(R_+,R_-) \, {\bf U}^{-1} ,
\end{eqnarray}
where
\begin{eqnarray}
R_\pm=\frac{k_{0z}-\kappa_\pm}{k_{0z}+\kappa_\pm}=e^{-2 i \phi_\pm},
\end{eqnarray}
and $\kappa_\pm \in \mathbb{C}$ are the eigenvalues of $\boldsymbol{\kappa}$. Since the eigenvalues of a unitary operator have modulus 1, this implies that $\kappa_\pm$ are purely imaginary.

\section{Elliptic Wave Packet derivation of Geometric and Phase-derived GH-Shifts}

The incoming wave packet, Eq. \eqref{momentum wf exact}, written in longitudinal and transverse coordinates is
\begin{eqnarray}\hspace*{-0.5cm}
    \Psi^{({\sf i})}(\bk)= \mathcal{N}  \  
    e^{-\frac{\Delta_l^2}{2}\left(k_l-k_{0l}\right)^2
    -\frac{\Delta_{t1}^2}{2} k_{t1}^2 -\frac{\Delta_{t2}^2}{2} k_{t2}^2 -i \bk\cdot \br_c} ,
    \label{incoming wf lt coord}
\end{eqnarray}
where $\mathcal{N} = \frac{\sqrt{\Delta_l\Delta_{t1} \Delta_{t2}}}{\pi^{3/4}}$. Using the general eigen reflection coefficient for each $\bk$ and spin state from the formalism developed in previous Section \ref{appedix reflection spin}, we can obtain the reflection wave function for each spin state. Here, for simplicity,  we assume the scalar case but the result can be straightforwardly generalized to the spinorial case. The reflected wave function is given in Eq.~\eqref{outgoing wf entangled exact}. Because this integral does not admit a closed form solution, in general, we proceed with the expansion and quasi-spherical approximation scheme of Eqs.~(\ref{ln r expansion}) and \eqref{elliptical kz 0th}. 

Under this approximation the resulting reflected wave function,
$\Psi^{({\sf r})}(\br,t)\approx\prod_{\gamma=x,y,z} \Psi^{({\sf r})}_\gamma(r_\gamma,t)$, becomes separable in terms of one-dimensional Gaussians, 
\begin{eqnarray}
    \Psi^{({\sf r})}_\gamma(r_\gamma,t)= {\cal N}_\gamma(t) \ e^{ - \frac{(r_\gamma - r^{({\sf r})}_{\gamma}(t))^2}{2 \Delta^2_\gamma(t)} + i \phi_\gamma(r_\gamma,t) } ,
\end{eqnarray}
with $\Psi^{(\sf r)}_\gamma = \Psi^{(\sf i)}_\gamma$ when $\gamma=x,y$ ($r_\gamma=\gamma$), centered at
\begin{eqnarray}
    r^{({\sf r})}_{x}(t) &=& x_c  + \frac{k_{0x}}{m} t , \quad
    r^{({\sf r})}_{y}(t) = y_c  , \nonumber \\
    r^{({\sf r})}_{z}(t) &=& - z_c +2 \tilde \delta_{\sf p} - \frac{\tilde k_{0z}}{m}t ,
\label{outgoing center scalar}
\end{eqnarray}
with dispersion $\Delta_\gamma(t) = \pi^{-\frac{1}{2}}/ |\mathcal{N}_\gamma (t)|^2$ expressed in terms of its normalization
\begin{eqnarray}
    \mathcal{N}_x (t)&=& \sqrt{\frac{\pi^{-\frac{1}{2}}\Delta_x}{ \Delta_x^2 + i\frac{t}{m} }} ,\quad 
    \mathcal{N}_y (t)= \sqrt{\frac{\pi^{-\frac{1}{2}}\Delta_y}{ \Delta_y^2 + i\frac{t}{m} }} \nonumber  \\ 
    \mathcal{N}_z (t)&=& \sqrt{\frac{\pi^{-\frac{1}{2}} \Delta_z}{ \Delta_z^2 + i\frac{t-2m\widetilde W^2}{m}}} \ e^{i (\widetilde\Phi -2 \tilde k_{0z} \tilde\delta_{\sf p})} .
\end{eqnarray}
Exact expressions for $\phi_\gamma$ are combersome, with asymptotic limits, $\lim_{t\rightarrow \infty} \phi_\gamma(r_\gamma,t)\rightarrow \tilde \phi_\gamma$,
\begin{eqnarray}
\label{phases scalar}
    \tilde \phi_x(x,t) &=& - \frac{k_{0x}^2 }{2m}t + k_{0x} (x - x_c) , 
    \quad \tilde \phi_y(y,t) = 0 ,\nonumber \\  
    \tilde  \phi_z(z,t) &=& - \frac{\tilde k_{0z}^2 }{2m}t - \tilde k_{0z} (z + z_c - 2 \tilde \delta_{\sf p}) ,
\end{eqnarray}
where only terms to $\mathcal{O}(1/t)$ are kept and we have made use of $k_{0y}=0$.

Since the GH-shift is the longitudinal distance between incoming and  outgoing reflection points, one can heuristically use $\Psi^{(\sf r)}_z$ to extract the geometric GH-shift $\lambda_x$. Note that $\Psi_z^{(\sf r)}$'s center is shifted by twice the penetration depth $2\tilde \delta_{\sf p}$, the factor of $2$ resulting from the identification of the classical path of an incoming particle traveling into the material before leaving the surface. While this interpretation yields illuminating insights for the GH-shift, it is only a semi-classical one. Particularly, $\Psi_{z}^{(\sf r)}$ is only the solution to the Schr\"odinger equation for $z<0$, thus it does not describe the motion of the incoming wave packet inside the material.  To obtain $\lambda_x$ one needs to 
compute the time interval 
\begin{align}
    \tau_{\rm WP} &= \frac{2\tilde\delta_{\sf p}}{v_z} = \frac{2 \tilde \delta_{\sf p}}{\tilde k_{0z} / m} ,
\end{align}
which yields Eq.~\eqref{gh shift elliptical}. Note that the second-order correction $\widetilde W^2$ does not affect $\lambda_x$. To measure this shift one would need a position-sensitive detector able to resolve such shift. 

Alternatively, one can  determine the relative phase-shift of the reflected wave packet and use the generalized ACH formula. The phase shift accumulated by the reflected wave function  $\Psi^{({\sf r})}(\br,t)$, Eq.~\eqref{elliptical spatial outgoing 0th entangled}, in the limit $t \rightarrow \infty$ is the sum of all phases in Eq.~\eqref{phases scalar}. The overall accumulated phase can be easily obtained by realizing that the normalization factor in this limit becomes
\begin{align}
    \lim_{t\rightarrow \infty} \mathcal{N} (t) \rightarrow \sqrt{ \frac{i m^3 \Delta_x \Delta_y \Delta_z }{\pi^{\frac{3}{2}} t^2 (t-2m\widetilde W^2) } } e^{i( \tilde \Phi - 2\tilde k_{0z} \tilde \delta_{\sf p})} ,
\end{align}
meaning that it contributes the non-trivial phase $\widetilde \Phi-2\tilde k_{0z} \tilde \delta_{\sf p}$. For unentangled particles the total reflection phase shift of the incoming wave packet is given by
\begin{align}
    \label{phase 2nd order}
    \lim_{t\rightarrow \infty} \left[\operatorname{Arg} ( \Psi^{({\sf r})}(\br,t)) -\tilde \bk^{(\sf r)}_0 \cdot \br + \frac{\tilde k_0^{({\sf r})2}}{2 m} t  \right]= \widetilde{\Phi}(\tilde k_{0z}),
\end{align} 
where we subtracted the kinetic phase. Using this wave packet analog of the reflection phase shift with the modified ACH formula, and noting that
\begin{eqnarray}
     \frac{d\widetilde \Phi(\tilde k_{0z})}{dk_{0z}} = \frac{d\widetilde \Phi(\tilde k_{0z})}{d\tilde k_{0z}} \  \frac{d\tilde k_{0z}}{dk_{0z}} = 2 \tilde \delta_{\sf p} \left( \frac{\Delta_l}{\Delta_z} \right)^2
\end{eqnarray}
one arrives the GH-shift in Eq.~\eqref{ACH formula correction} for the scalar case. Comparison between ACH and geometric GH-shifts for different incoming states can be found in Table~\ref{TableI}. 

We also remark that the $\widetilde W^2$-correction does not affect the GH-shift. Rather,  $\widetilde{W}^2$ modifies the effective wave packet dispersion. For beams reflected from a semi-infinite slab with a constant optical potential $\widetilde W^2 = \tilde k_{0z} \tilde \delta_{\sf p}^3$, this correction is relevant for precision-measurement experiments involving mirrors (as in Mach-Zehnder interferometers) that have large penetration depths $\tilde \delta_{\sf p}$. If the slab is magnetic, as we will see, its effect will be different on each spin component of a neutron wave packet, thus affecting the measurement contrast at the detector. 

\begin{table}[t]
\centering
\renewcommand{\arraystretch}{2}
\begin{tabular}{|c|c|c|}
    \hline
     & $\lambda^{\rm ACH}_x$ ($t\rightarrow\infty$) & $\lambda_x$   \\
     \hline
    PW & $2\delta_{\sf p} \cot\alpha$ & N/A \\
    SWP & $2\delta_{\sf p} \cot\alpha$ & $2\delta_{\sf p} \cot\alpha$  \\
    EWP & $\left (\frac{\Delta_l}{\Delta_z}\right)^2 2\tilde \delta_{\sf p}  \cot \alpha $  & $\left (\frac{\Delta_z}{\Delta_l}\right)^2 2\tilde \delta_{\sf p}  \cot \alpha $  \\
    \hline
\end{tabular}
\caption{GH-shifts for different incoming states (PW: plane wave, SWP: Spherical wave packet and EWP: Elliptical wave packet). The 
ACH formula, $\lambda^{\rm ACH}_x$, is derived from 
the asymptotic reflection phase-shift $\widetilde{\Phi}(\tilde k_{0z})$ measured at time $t\rightarrow\infty$. }
\label{TableI}
\end{table}

If the beam carries spin (or polarization) and the surface's optical potential couples to it, then each surface eigen spinor will generally experience different penetration depths, $\tilde \delta_{{\sf p}\pm}$,  and thus, ultimately, different GH-shifts. Here, we use the subscript $\nu=\pm$ to denote the corresponding surface eigen spin state's quantities, i.e.  $\tilde \delta_{{\sf p}\nu}$, $\lambda_{x\nu}$, $\widetilde \Phi_\nu$ and $\widetilde W^2_\nu$. Then,  $\lambda^{\rm ACH}_{x+}-\lambda^{\rm ACH}_{x-}$ can be observed by extracting the phase-shift difference in a reflected entangled beam from polarization measurements and using the generalized ACH formula, Eq.~\eqref{eq:GACH}. Carrying out approximations similar to the un-polarized case above, one can factorize $\Psi^{({\sf r})}_{\mu\nu}(\br,t)\approx\prod_{\gamma=x,y,z} \Psi^{({\sf r})}_{\gamma\mu\nu}(r_\gamma,t)$, with 
$\Psi^{({\sf r})}_{\mu\nu}(\br,t)$ as  defined in Eq.~\eqref{elliptical spatial outgoing 0th entangled}, and where 
the centers  are modified due to the spin-path entanglement and the spin-dependent penetration depths, i.e., 
\begin{eqnarray}
    \nonumber
    r^{(\sf r)}_{x\mu} (t) &=& r^{(\sf r)}_x(t) + \mu \frac{\xi_x}{2}, \quad  r^{(\sf r)}_{y\mu} (t) = r^{(\sf r)}_y(t) +\mu \frac{\xi_y}{2} ,\\  
    r^{(\sf r)}_{z\mu\nu}(t) &=&  - z_c +2 \tilde \delta_{{\sf p}\nu} - \frac{\tilde k_{0z}}{m}t - \mu \frac{\xi_z}{2}.
    \label{outgoing center}
\end{eqnarray}
Similarly, asymptotic phases $\tilde \phi_{\gamma\mu\nu}(r_\gamma,t)$ become 
\begin{eqnarray} 
&&\hspace*{-0.4cm} \tilde \phi_{x\mu\nu}(x,t) = \tilde \phi_x(x,t) - \mu  \frac{k_{0x} \xi_x}{2}  , 
    \quad \tilde \phi_{y\mu\nu}(y,t) =  0 ,\nonumber \\  
    &&\hspace*{-0.4cm} \tilde  \phi_{z\mu\nu}(z,t) = \tilde  \phi_{z}(z,t) + 2\tilde k_{0z} ( \tilde \delta_{{\sf p}\nu}-\tilde \delta_{\sf p}) + \mu \frac{\tilde k_{0z}\xi_z}{2}.
    \label{phases spinor}
\end{eqnarray}

On the other hand, the dispersion $\Delta_{\gamma\nu}(t)=\pi^{-\frac{1}{2}}/|\mathcal{N}_{\gamma\nu}(t)|^2$ and normalization $\mathcal{N}_{\gamma\nu}$ are $\mu$-independent
\begin{eqnarray}
    \nonumber
    \mathcal{N}_{x\nu}(t) &=& \mathcal{N}_{x}(t), \quad \mathcal{N}_{y\nu}(t)= \mathcal{N}_{y}(t) , \\
    \mathcal{N}_{z\nu} (t)&=& \sqrt{\frac{\pi^{-\frac{1}{2}} \Delta_z}{ \Delta_z^2 + i\frac{t_\nu}{m}}} \ e^{i (\widetilde\Phi_\nu -2 \tilde k_{0z} \tilde\delta_{{\sf p}\nu})} ,
\end{eqnarray}
where $t_\nu=t-2m\widetilde W_\nu^2$. Also,  we see that $ \Psi^{(\sf r)}_{\gamma\mu\nu } $ is indeed $\nu$-independent when $\gamma=x,y$, i.e.,
\begin{eqnarray}
 \Psi^{(\sf r)}_{\gamma\mu\nu }= \Psi^{(\sf i)}_{\gamma\mu}   .
 \label{xy property}
\end{eqnarray}
This property is a simple generalization of the scalar case. To compute the polarization we assume the detector is a volume-integrated one
\begin{eqnarray}
    P_\gamma(t) = \int d\br \ \Psi^{(\sf r)\dagger}_\bxi(\br,t) \ \sigma^\gamma \ \Psi^{(\sf r)}_\bxi(\br,t), 
    \label{polarization integral}
\end{eqnarray}
leading to Eq.~\eqref{polarization master} that we evaluate next. 

In the $t\rightarrow \infty$ limit, there is a significant simplification  
\begin{eqnarray}
\nonumber
    \Psi^{(\sf r)*}_{\mu\nu} (\br,t) \Psi^{(\sf r)}_{\mu'\nu'} (\br,t) \xrightarrow{t\rightarrow\infty} |\Psi^{(\sf r)*}_{\mu\nu} (\br,t) \Psi^{(\sf r)}_{\mu'\nu'} (\br,t) |  e^{i \eta_{\mu\mu'\nu\nu'}}, \\
\end{eqnarray}
where $\eta_{\mu\mu'\nu\nu'}$ becomes $\br$-independent. Therefore, the integral, which encodes the overlap between two Gaussians, can be evaluated in closed form 
\begin{eqnarray}
\label{amplitude}
\nonumber
& &\Lambda_{\mu\mu'\nu\nu'} =  \Lambda_{\nu\nu'} \, e^{- \frac{m^2(\mu - \mu')^2}{8} \Big ( \frac{\Delta_x^2 \xi^2_{x}+\Delta_y^2 \xi^{2}_{y}}{2t^2}+\frac{\Delta_z^2 \xi^{(\sf r)^2}_{z,\mu\mu'\nu\nu'}} {t_\nu^2 + t_{\nu'}^2} \Big )} ,
\end{eqnarray}
with $\Lambda_{\nu\nu'}=\sqrt{\frac{2 t_\nu \, t_{\nu'}}{ t_\nu^2 + t_{\nu'}^2 }}$, $\xi^{(\sf r)}_{z,\mu\mu'\nu\nu'}=-\xi_{z} + 4 \frac{\tilde\delta_{{\sf p}\nu} -  \tilde \delta_{{\sf p}\nu'}}{\mu-\mu'}$. (When the magnetic field aligns with the incoming entangled basis $\bxi^{(\sf r)}_\|=(\xi_x,\xi_y)$ and $\xi^{(\sf r)}_z = \frac{\mu-\mu'}{2} \xi^{(\sf r)}_{z,+-+-}$ reduces to
\begin{eqnarray}\hspace*{-0.7cm}
\xi_z^{({\sf r})} = -\xi_z+2(\tilde \delta_{{\sf p}+} - \tilde \delta_{{\sf p}-}) .
\label{outgoing xi entangled}
\end{eqnarray}

The case of an unentangled incoming beam  provides the simplest illustration. Consider a particle beam polarized along the ${x}$ direction (this corresponds to putting $\bxi= 0$ in $\mathbf{\chi}^{(\sf i)}_{\bk\cdot\bxi}$). Then, 
%
%
\begin{eqnarray} {\hspace*{-0.7cm}}
    \label{polarization simplified2}
    P_x &=& \sin^2\theta  \cos^2 \phi + A (1-  \sin^2\theta \cos^2\phi)  \cos\Delta \widetilde \Phi_{+-}  \\
    P_y &=& -A\cos\theta \sin\Delta\widetilde\Phi_{+-} + \frac{1- A \cos\Delta\widetilde\Phi_{+-}}{2} \sin^2\theta  \sin2\phi   \nonumber\\
    P_z &=&  \sin\theta (  \cos\theta \cos \phi \ (1-A \cos \Delta \widetilde \Phi_{+-} ) \nonumber \\ 
    &&\hspace*{3cm}+ A \sin\phi \sin\Delta \widetilde \Phi_{+-} ) \nonumber,
\end{eqnarray}
where $\Delta \widetilde \Phi_{+-} = \eta_{\mu\mu'+-}$ and $A=\Lambda_{\mu\mu'+-}$
This polarization satisfies 
$P_x^2+P_y^2+P_z^2= \sin^2 \theta \cos^2 \phi  (1-A^2) +A^2  \le 1$.

The polarization in Eq.~\eqref{polarization integral} assumes that the reflected waves travel freely from the surface to the detector. However, when the incoming wave packet is $\Psi^{(\sf i)}_\bxi$ one typically positions a {\it disentangler}  before detection. We show next the theory behind the effect of the disentangler on the measured polarization (Sec. \ref{spin-echo-reflectometry} describes the mathematics involved in spin-echo reflectometry; for the experimental realization see Ref. \cite{SERexp}).  

Assume the particle beam is entangled with ${\cal U}_{\hat {\sf e}}^{{\sf se}}(\bxi)$, and the {\it disentangler}, ${\cal U}_{\hat {\sf d}}^{{\sf se}}(\bxi^{(\sf se)})$,  is set along the quantization axis ${\hat{\sf d}}=\hat n$ at time $t=t_{\rm se}$ with entanglement vector $\bxi^{(\sf se)}=-(\xi_x,\xi_y,-\xi_z)$. We note here that this setting of $\bxi^{(\sf se)}$ is particular to the case of reflectometry, in contrast to the usual transmission mode, where $\bxi^{(\sf se)}=-\bxi$. The reason behind this is that the reflection operator has the form 
\begin{eqnarray}
     \hat R(k_z)
    \ket{-k_z}\bra{k_z}
\end{eqnarray}
Hence, in order for the {\it disentangler} to be the desired inversion of the {\it entangler}, the $k_z$-transfer matrix part needs to be $\ket{k_z}\bra{-k_z}$, which  effectively inverts $\xi_z$.

The action of the disentangler transforms the reflected wave function as 
\begin{eqnarray}\hspace*{-0.5cm}
    \Psi_{\mu\nu}^{(\sf r)}(\br,t_{\sf se}) \rightarrow \Psi_{\mu\nu}^{(\rm se)}(\br,t_{\sf se}) = \Psi_{\mu\nu}^{(\sf r)}(\br +\nu \frac{\bxi^{(\sf se)}}{2},t_{\sf se}). 
\end{eqnarray}
Due to this shift, after leaving the {\it disentangler} $\Psi^{(\rm se)}_{\mu\nu}$ has its center and phase, 
from Eq.~\eqref{outgoing center} and \eqref{phases spinor}, modified to
\begin{eqnarray}
    \nonumber
    r^{(\sf se)}_{x\mu\nu} (t) &=& r^{(\sf r)}_{x\mu}(t) -\nu \frac{\xi_x}{2}, \quad  r^{(\sf se)}_{y\mu\nu} (t) = r^{(\sf r)}_{y\mu}(t) -\nu \frac{\xi_y}{2} ,\\  
    r^{(\sf se)}_{z\mu\nu}(t) &=&  r^{(\sf r)}_{z\mu\nu}(t) +
    \nu \frac{\xi_z}{2} ,
    \label{spin echo center}
\end{eqnarray}
\begin{eqnarray} 
&&\hspace*{-0.4cm} \tilde \phi^{(\sf se)}_{x\mu\nu}(x,t) = \tilde \phi_{x\mu\nu}(x,t)+\nu
\frac{k_{0x} \xi_x}{2}  , 
    \quad \tilde \phi^{(\sf se)}_{y\mu\nu}(y,t) =  0 ,\nonumber \\  
    &&\hspace*{-0.4cm} \tilde  \phi^{(\sf se)}_{z\mu\nu}(z,t) = \tilde  \phi_{z\mu\nu}(z,t) - \nu \frac{\tilde k_{0z}\xi_z}{2}.
    \label{spin echo phases spinor}
\end{eqnarray}
The polarization $P_\gamma$ can now be computed and the result is of the form of Eq.~\eqref{polarization master} with the modification
\begin{eqnarray}
    \Lambda^{(\sf se)}_{\mu\mu'\nu\nu'} &=&  \Lambda_{\nu\nu'} \, e^{- \frac{m^2(\mu - \mu'+\nu'-\nu)^2}{8} \Big ( \frac{\Delta_x^2 \xi^2_{x}+\Delta_y^2 \xi^{2}_{y}}{2t^2}+\frac{\Delta_z^2 \xi^{(\sf se)^2}_{z,\mu\mu'\nu\nu'}} {t_\nu^2 + t_{\nu'}^2} \Big )} , \nonumber \\
\eta^{(\sf se)}_{\mu\mu'\nu\nu'} &=& \eta_{\mu\mu'\nu\nu'} + \frac{\nu'-\nu}{2} \sf \widetilde\bk_0 \cdot \bxi \ , 
\end{eqnarray}
where $\xi^{(\sf r)}_{z,\mu\mu'\nu\nu'}=-\xi_{z} + 4 \frac{\tilde\delta_{{\sf p}\nu} -  \tilde \delta_{{\sf p}\nu'}}{\mu-\mu'+\nu'-\nu}$. We see that only the amplitude terms $\Lambda^{(\sf se)}_{++--}$ and $\Lambda^{(\sf se)}_{--++}$ are $\bxi$-independent. If the exponential damping is substantial, this means that these two terms will be the main contribution to $P_\gamma$. Interestingly, the phase difference $\eta^{(\sf se)}_{++--}$ and $\eta^{(\sf se)}_{--++}$ are also $\bxi$-independent. Therefore, $P_\gamma$ might appear to not contain $\bxi$, especially if one normalizes the $P_\gamma$ by purely the spin-analyzer's count, instead of measuring the reflectivity. The reflectivity because it does not have cross-terms from different reflected spin-states.

Note that in those cases where ${\hat{\sf e}}={\hat{\sf d}}=\hat n$, i.e., the perfect spin-echo situation, the measured polarization does not depend on $\bxi$.

Different from the previous cases, when the quantization directions  ${\hat{\sf e}}={\hat{\sf d}}\ne \hat n$, the action of the {\it disentagler} on each $\Psi_{\mu\nu}^{(\sf r)}$ is now 
\begin{eqnarray}
   \Psi_{\mu\nu}^{(\sf r)}(\br,t_{\sf se}) \ket{\nu} &\rightarrow&  |\Psi_{\mu\nu}^{(\rm se)}(t_{\sf se}) \rangle \\
   \nonumber
   & =&  \ \Psi_{\mu\nu}^{(\sf r)}(\br + \frac{\bxi^{(\sf se)}}{2},t_{\sf se})  _{\hat{\sf d}}\braket{+}
   {\nu}_{\hat n}  \ket{+}_{\hat{\sf d}}\\
    \nonumber
   & &+\Psi_{\mu\nu}^{(\sf r)}(\br - \frac{\bxi^{(\sf se)}}{2},t_{\sf se})  _{\hat{\sf d}}\braket{-}
   {\nu}_{\hat n}  \ket{-}_{\hat{\sf d}}
\end{eqnarray}
Now, the action of the {\it disentangler} can no longer be understood purely in terms of the wave packet center and phase shift as before. As an example, we consider explicitly the case $\hat{\sf e}={\hat{\sf d}}=\hat z$ and $\hat n=\hat 
y$, while the exact polarization can be calculated in closed form, the expression is cumbersome and un-enlightening, thus we give here the result in the incoming plane-wave limit
\begin{eqnarray}
    P_x &=& \cos^2 \bk_0 \cdot \bxi \ \cos\Delta  \Phi_{+-} \   +
    \sin^2  \bk_0 \cdot \bxi , \nonumber\\
    P_y &=& \sin2\bk_0 \cdot \bxi \left (\frac{1- \cos\Delta\Phi_{+-}}{2}\right ) , \\
    P_z &=&  \cos \bk_0 \cdot \bxi \ \sin \Delta  \Phi_{+-}  ,    \nonumber
    \label{polarization spin echoed}
\end{eqnarray} 
where $P_x^2+P_y^2+P_z^2=1$ and all the information about the layered structure is contained in $\Delta \Phi_{+-}$. For practical applications with neutron beams, we need to consider 2 important experimental limitations. The accessible entanglement lengths $\xi$ are in the range of $35\mu m$, and the beam divergence is typically of the order of $10$ mrad. Hence, the polarization is highly oscillating. When plotted as a function of $k_z$, one can average (or coarse- grain) over a window including several of these oscillations leading to the effective observed polarization
\begin{eqnarray}
\nonumber
    P^{(\sf obs)}_x &=& \frac{1+\cos \Delta\Phi_{+-}}{2}, \\  
    P^{(\sf obs)}_y&=&P^{(\sf obs)}_z=0.
\end{eqnarray}
We also observe that $P_x^{(\sf obs)}$ to have higher sensitivity to the magnetic and non-magnetic densities when the probed layer is contained in a specially-engineered multi-layer sample to have large GH-shifts. This suggests that entangled reflectometry in TR can be utilized as complementary to the usual polarized reflectometry.

\section{Probing Non-Magnetic and Magnetic in-plane Structures}
\label{appendix surface}

We are interested in probing non-magnetic and magnetic structures that arise on the surface of a material slab or thin film by exploiting entanglement in the quantum probe. 
The  approach we describe next differs from the \textit{phase-object approximation} \cite{phaseobject1,phaseobject2,phaseobject3}. To keep notation compact, hereinafter we assume  the origin of time, $t=0$, when $\Psi^{(\sf r)}_\bxi$'s centers $\br^{(\sf r)}_{\mu\nu}(t)$, Eq.~\eqref{outgoing center}, leave the surface. It is important to note that if the penetration depth difference $|\tilde \delta_{{\sf p}+}-\tilde\delta_{{\sf p}-}|$ is large, or $\bxi$ has a non-zero $z$-component, the centers $\br^{(\sf r)}_{\mu\nu}(t)$ will leave the surface at different times. This differs from the usual phase object approximation, but since measurements are performed in the asymptotic limit $t\rightarrow\infty$, such initial temporal delay should not affect the final outcome. Nonetheless, the initial spatial separation between incoming coherent spinors, $\bxi$, and the outgoing additional separation due to GH-shift, are still taken into account in our formalism. These separations show up in both amplitude $\Lambda_{\mu\mu'\nu\nu'}$ and phase $\eta_{\mu\mu'\nu\nu'}$ (Eq.~\eqref{outgoing phase diff}) at the detector.

Our approach starts with the reflected state 
%
\begin{eqnarray}
 \Psi^{(\sf r)}_{\bxi} (\bk)= \bra{\bk} \hat {\cal R} \,|\Psi^{(\sf i)}_{\bxi}\rangle , 
 \label{eq:localapprox}
\end{eqnarray}
where the unitary operator is given by
\begin{eqnarray}\hspace*{-0.6cm}
    \hat{\cal R} = \!\!\int \! d\bk  \!\!\int \! \frac{d\br_\|}{2\pi} \hat{R} (\br_\|,k_z)  e^{ -i \bk_\| \cdot \br_\|}\! \ketbra{\bk_\|}{\br_\|}\otimes\ketbra{-k_z}{k_z}  , 
    \label{eq:Roperator}
\end{eqnarray}
and $\hat R(\br_\|,k_z)$ is the reflection operator derived in Sec.~\ref{appedix reflection spin}. Then, its time-evolved, reflected wave packet becomes
\begin{eqnarray}
    \hspace*{-0.5cm}
    \Psi^{(\sf r)}_{\bxi} (\br,t) &=& \frac{1}{(2\pi)^{\frac{3}{2}}} \int d\bk \  \Psi^{(\sf r)}_{\bxi} (\bk) \ e^{i (\bk^{(\sf r)} \cdot \br - \frac{k^{(\sf r)2}}{2 m} t)} .     \label{outgoing phase object}
\end{eqnarray}

The particular case of non-magnetic samples where only inhomogeneities in the nuclear density are present in the sample, i.e., $\hat{R}(\br_\|, k_z) = R(\br_\|,k_z) \mathbb{1}$, can be easily analyzed. The reflected wave function can be re-expressed 
\begin{eqnarray} 
\Psi_\bxi^{(\sf r)}(\bk,t) &=& 
 \frac{1}{\sqrt{2}} \sum_{\mu=\pm} \int \frac{d\br_\|}{2\pi}  \ e^{-i \bk_\| \cdot \br_\|} \nonumber \\
& &\hspace*{-0.5cm} \times R(\br_\|,k_z) \Psi^{(\sf i)}(\br_{\|\mu},k_z,0) e^{-i\frac{k^{(\sf r)2}}{2m}t}\ket{\mu}_{\hat{\sf e}} , 
\label{wf reflected explicit}
\end{eqnarray}
involving only the index $\mu$ because $\hat{R}(\br_\|,k_z)$ is diagonal, $\br_{\|\mu}=\br_\|-\mu \frac{\bxi}{2}$, and
\begin{eqnarray}\hspace*{-0.5cm}
\Psi^{(\sf i)}(\br_{\|\mu},k_z,t)=  \int \frac{d\bk_\|}{2\pi} \, e^{i(\bk_\| \cdot \br_{\|\mu}-\frac{k^{(\sf r)2}}{2m}t)}   \Psi^{(\sf i)}(\bk_\|,k_z)  .
\end{eqnarray}

To proceed, we assume that the entanglement vector is confined to the $xy$-plane, i.e., $\bxi=\bxi_\|$, and a spin-echo measurement is performed where the disentangler's settings are $\bxi^{(\sf se)} = -\bxi_\|$ and $\hat{\sf d}=\hat{\sf e}$. While details of spin-echo reflectometry are given in the next section, we need the disentangler operator in Eq.~\eqref{spin echo momentum} to write the state after the disentangler
\begin{eqnarray}
    \Psi^{(\sf se)}_\bxi(\br_\|,k_z,t)  &=& \langle \br_\|,k_z |  \mathcal{U}^{\sf se}_{\hat{\sf e}}(\bxi^{({\sf se})}) | \Psi^{(\sf r)}_\bxi(t)\rangle \\
    \nonumber
    && \hspace*{-1cm}=\frac{1}{\sqrt{2}}\sum_{\mu=\pm} \int d\br'_\| \ \delta^2(\br_\| +\mu \frac{\bxi_\|}{2} - \br'_\|) \\
    \nonumber
    & &\times R(\br'_\|,k_z) \Psi^{(\sf i)}(\br'_{\|\mu},k_z,t)  \ket{\mu}_{\hat{\sf e}}\\
    \nonumber
    && \hspace*{-1cm}=\frac{1}{\sqrt{2}} \sum_{\mu=\pm}  R(\br_\| + \mu \frac{\bxi_\|}{2},k_z) \Psi^{(\sf i)}(\br_\|,k_z,t) \ket{\mu}_{\hat{\sf e}} .
\end{eqnarray}
One can then compute, e.g., the polarization \begin{eqnarray}
    \nonumber
    P_\gamma(\bxi_\|,t) &=& \int d\br_\| dk_z \Psi^{(\sf se)\dagger}_\bxi (\br_\|,k_z,t) \sigma^\gamma  \Psi^{(\sf se)}_\bxi (\br_\|,k_z,t) ,
\end{eqnarray}
$\gamma=x,y,z$, which after averaging over the impact parameters $(x_c,y_c)$  yields Eq.~\eqref{grating-reflection} for $\gamma=x$.

Next, we consider the situation where in-plane inhomogeneities are magnetic. In this case, since $\hat n(\br_\|)$ varies over the surface, we assume that the sample's reflection operator does not change appreciably over a distance $\Delta_\gamma$ 
\begin{eqnarray}
    |\hat{R}(\br_\|, \tilde k_{0z})| &\gg& {\Delta_\gamma} \left|\frac{\partial \hat R (\br_\|, \tilde k_{0z}) }{\partial\br_\gamma} \right|   , \quad \gamma=x,y ,
    \label{approximation condition}
\end{eqnarray}
a condition that allows us to further approximate $\Psi^{(\sf r)}_{\bxi} (\bk)$ by making the following replacement in Eq. \eqref{eq:localapprox} 
\begin{eqnarray}
\hat{R} (\br_\|,k_z) \ket{\mu}_{\hat{\sf e}} \rightarrow \hat{R} (r^{({\sf r})}_{\gamma \mu}(t=0),k_z) \ket{\mu}_{\hat{\sf e}} .
\end{eqnarray}
Note that this replacement does not preserve unitarity. Therefore, one needs to normalize $\Psi^{(\sf r)}_{\bxi} (\br,t)$ in Eq. \eqref{outgoing phase object}. In addition, if the incoming wave packet is elliptical we use the quasi-spherical approximation of Eq. \eqref{elliptical kz 0th} (leading to a separable reflected wave packet).

The polarization of the reflected wave packet with initial impact parameter $\br_c$, $P_\gamma (\br_{c},\bxi)$, is calculated by substituting Eq.~\eqref{outgoing phase object} into \eqref{polarization integral}. The result has the form of Eq.~\eqref{polarization master} but with quantities, $\bU_{\mu\nu}, \widetilde\Phi_{\nu}, \tilde\delta_{{\sf p}\nu}$, and $\widetilde W^2_{\nu}$, spatially dependent on $r^{(\sf r)}_{\gamma\mu}(t=0)$. The observed beam polarization is averaged over the area defined by the impact parameter, $\br_{c}$, assumed bigger than that of the characteristic unit cell, $p$, of the magnetic (or birenfringet) surface structure to be probed 
\begin{eqnarray}
      \overline{P}_{\gamma}(\bxi) &=& \frac{1}{A} \int_A dx_c d y_c   \ P_\gamma (\br_{c},\bxi) ,
\end{eqnarray}
which satisfies strictly
\begin{eqnarray}
  \overline{P}_x^2(\bxi)+\overline{P}_y^2(\bxi)+\overline{P}_x^2(\bxi) \le 1  ,
\end{eqnarray}
and it sums to 1 when $\bxi={\bf 0}$.

As an illustration consider the case of incoming entangled wave packets spatially separated by $\bxi=\xi \hat{x}$ with spin polarization $\ket{\pm}_{\hat z}$. The centers of those wave packets experience different magnetic fields ${\bf B}(r^{(\sf r)}_{\gamma+}(t=0) )=B \hat{x}$ and ${\bf B}(r^{(\sf r)}_{\gamma-}(t=0))=B \hat{y}$, implying $\bU (r^{(\sf r)}_{\gamma+}(t=0)) = \hat{U}_x$ and $\bU (r^{(\sf r)}_{\gamma-}(t=0) ) = \hat{U}_y$. We further assume that $\Delta_\gamma\rightarrow0$ for the sake of clarity. 
The resulting normalization of the reflected state $\Psi^{(\sf r)}_{\bxi} (\br,t)$ is 
\begin{eqnarray}
   \lim_{\Delta_\gamma \rightarrow 0}|\mathcal{N}(\br_{c},\bxi)|^2&& \\
   \nonumber
   && \hspace*{-2.2cm}= \frac{1}{2} \left[ 2 + \cos (k_{0x} \xi) + \sqrt{2} \sin (k_{0x} \xi) \sin \left( \Delta \widetilde\Phi_{+-} + \frac{\pi}{4} \right)  \right],
\end{eqnarray}
which can be greater or less than $1$ depending on  $k_{0x}\xi$.

\section{Spin-Echo Reflectometry}
\label{spin-echo-reflectometry}

In this section we discuss the effect of a spin-echo measurement on the observed beam polarization. See Fig.~\eqref{setup-entangled} for a setup. 

The entangled wave packet of Eq.~\eqref{momentum wf entangled} is generated by an {\it entangler}, an instrument such as a pair of Magnetic Wollaston prisms or RF-flippers \cite{shufan}. Operationally, the entangler can be represented as a unitary operator
\begin{eqnarray}\hspace*{-0.6cm}
    \mathcal{U}^{\sf se}_{\hat{{\sf e}}}(\bxi) &=& \!\!\int \!d\bk  \ketbra{\bk}{\bk} \otimes (e^{-i \frac{\bk \cdot \bxi}{2}} |+\rangle_{\!{\hat{\sf e}}} \!\langle +|+  e^{i \frac{\bk \cdot \bxi}{2}}|-\rangle_{\!{\hat{\sf e}}} \!\langle -| ),  \label{spin echo momentum}
\end{eqnarray}
in momentum basis, or equivalently as
\begin{eqnarray}\hspace*{-0.6cm}
    \mathcal{U}^{\sf se}_{\hat{{\sf e}}}(\bxi) &=& \!\!\int \! d\br ( |\br_+ \rangle  \langle \br| \otimes |+\rangle_{\!{\hat{\sf e}}} \!\langle +| + |\br_- \rangle  \langle \br| \otimes |-\rangle_{\!{\hat{\sf e}}} \!\langle -|) ,     
\label{spin echo position} 
\end{eqnarray} 
in position basis, with $\br_\pm=\br \pm\frac{\bxi}{2}$, and $\ket{\pm}_{\hat{\sf e}}$ eigenstates of $\boldsymbol{\sigma}\cdot{\hat{\sf e}}$ with ${\hat{\sf e}}$ an arbitrary quantization axis; Eq. \eqref{momentum wf entangled} corresponds to ${\hat{\sf e}}=\hat z$. In transmission mode the {\it disentangler}, which reverses the phase difference and separation of spinor components introduced by the entangler,  corresponds to $\mathcal{U}^{\sf se}_{\hat{\sf e}}(\bxi)^\dagger = \mathcal{U}^{\sf se}_{\hat{\sf e}}(-\bxi)$. 

To understand the action of $\mathcal{U}^{\sf se}_{\hat{\sf e}}(\bxi)$ on an input state, consider  a polarized plane wave $\ket{\bk} \otimes \ket{+}_{\hat x}$. Then, the action of $\mathcal{U}^{\sf se}_{\hat z}(\bxi)$ on that state generates the spinor $\boldsymbol{\chi}^{({\sf i})}_{\bk\cdot \bxi}$ of Eq. \eqref{momentum wf entangled}. Similarly, if the input state is $\ket{\br}\otimes\ket{+}_{\hat x}$, $\mathcal{U}^{\sf se}_{\hat z}(\bxi)$ shifts $\ket{\pm}_{\hat z}$'s initial positions $\ket{\br}$ to $\ket{\br_\pm}$. Hence, because a polarized wave packet can be written as a superposition of either $\bk$ or $\br$, it will experience both shifts. Particularly, the imparted phase becomes the phase $\mp\bk \cdot \bxi/2$, while the position shift becomes its center shift $\br_c \pm \bxi/2$. Due to the displacement of the wave packet center, if one simply performs a measurement without using the disentangler, $\mathcal{U}^{\sf se}_{\hat z}(-\bxi)$, the observed polarization will generally display an exponential damping term, as $|\bxi|$ increases, due to the lack of overlap between the two spinor components. If the interaction between the beam and the sample imparts a (momentum-dependent or independent) relative phase shift between those components without changing the quantization axes (i.e., $\hat n=\hat{\sf e}$), one can always set up the disentangler with ${\hat{\sf d}}=\hat{\sf e}$ in order to  cancel the $\bxi$-dependence in the measured polarization, thus uncovering the relative phase shift. The GH-shift is associated to a momentum-dependent situation.   

In the general case where the interaction with the sample also changes the incoming spinor basis $\ket{\pm}_{\hat{\sf e}}$, the disentangler cancellation logic above breaks down. In principle, one might attempt to change the disentangler basis (after reflection) to be the same as the sample's spinor basis, in our notation $\hat{\sf d} = \hat{n}$, but this is impractical for the case the basis change is inhomogeneous ($\hat{n} (\br_\|)$). 
In our entangled-beam reflectometry setup, the action of $\mathcal{U}^{\sf se}_{\hat{\sf d}}(\bxi^{(\sf se)})$ on our reflected wave packet $|\Psi^{(\sf r)}_{\bxi}\rangle$, at $t=t_{\sf se}$ with $\bxi^{(\sf se)}=-(\xi_x,\xi_y,-\xi_z)$, is
\begin{eqnarray}
    \langle \bk|  \mathcal{U}^{\sf se}_{\hat{\sf d}}(\bxi^{(\sf se)}) |\Psi_\bxi^{(\sf r) } (t_{\sf se}) \rangle = \langle \bk |\Psi^{({\sf se})}_\bxi (t_{\sf se}) \rangle  && \nonumber \\
 && \hspace*{-5cm}   =  \sum_{\mu,\nu} \bU_{\mu\nu}^* \Psi^{(\sf r)}_{\mu\nu}(\bk,t_{\sf se}) ( e^{-i \frac{\bk^{(\sf r)} \cdot \bxi^{(\sf se)}}{2}} \ {}_{\hat{\sf d}}\! \braket{+}{\nu}_{\hat{n} } \ket{+}_{\hat{\sf d}} \nonumber \\  
    && \hspace*{-3.0cm}+ e^{i \frac{\bk^{(\sf r)} \cdot \bxi^{(\sf se)}}{2}} \  {}_{\hat{\sf d}}\!\braket{-}{\nu}_{\hat n}  \ket{-}_{\hat{\sf d}} ),
    \label{spin echo wf d=n}
\end{eqnarray}
where $\Psi^{(\sf r)}_{\mu\nu}(\bk,t_{\sf se}) = 1/(2\pi)^\frac{3}{2} \int d\br \ e^{-i \bk\cdot \br} \Psi^{(\sf r)}_{\mu\nu}(\br,t_{\sf se})$ with $\Psi^{(\sf r)}_{\mu\nu}(\br,t)$ in Eq.~\eqref{elliptical spatial outgoing 0th entangled}. Let us consider first the case ${\hat{\sf d}}\ne\hat n$. Then, the inner-products $_{\hat{\sf d}}\!\braket{\pm}{\nu}_{\hat{n}}$ contain extra phases that will show up in the measured polarization. If, on the other hand ${\hat{\sf e}}\ne{\hat{\sf d}}=\hat n$, then 
the disentangler imparts the phase $\pm \bk^{(\sf r)} \cdot \bxi^{(\sf se)}/2 = \pm \bk \cdot \bxi $ to the state $\Psi^{(\sf r)}_{\mu\pm}$. However, from Eq.~\eqref{momentum wf entangled} the phase imparted on $\Psi^{(\sf r)}_{\mu\pm}$ by the original entangler is $-\mu \bk\cdot \bxi/2$. Hence, the  phases $\pm\bk\cdot \bxi/2$ cancel out in $\Psi^{(\sf r)}_{++}$ and $ \Psi^{(\sf r)}_{--}$.  Consequently, their wave packets' centers are also $\bxi$-independent but differ only by their corresponding GH-shifts. This is not the case  for $\Psi^{(\sf r)}_{-+}$ and $\Psi^{(\sf r)}_{+-}$, where their centers get displaced further apart because of the extra $\bxi$-dependence. While the analysis here focuses on the $\Psi^{(\sf se)}_{\mu\nu}$'s phases and centers, in order to assess the full impact of the entangler and disentangler on observables, such as reflectivity and polarization, these are not the only two relevant factors. Eq.\eqref{spin echo wf d=n} indicates that $\bU_{\mu\nu}$ affects the magnitude $|\Psi^{(\sf se)}_{\mu\nu}|$, thus certain components $\Psi^{(\sf se)}_{\mu\nu}$ might contribute more significantly to observables depending on the orientation of $\hat{\sf e}$ and $\hat{n}$.

\end{document}